\documentclass[aps,prl,reprint,superscriptaddress]{revtex4-1}  % linenumbers
\usepackage[dvipsnames]{xcolor}
\usepackage[colorlinks,allcolors={ForestGreen}]{hyperref}
\usepackage{graphicx}
\usepackage{color}
\usepackage{amsmath}
\newcommand{\ket}[1]{| #1 \rangle}
\newcommand{\bra}[1]{\langle #1 |}
\begin{document}
\title{\textit{Ab initio} optical potentials and nucleon scattering on medium mass nuclei}%

\author{A. Idini}
\affiliation{Department of Physics, University of Surrey, Guildford, GU2 7XH, UK} 
\affiliation{Division of Mathematical Physics, Department of Physics, LTH, Lund University, Post Office Box 118, S-22100 Lund, Sweden} 
\author{C. Barbieri}
\affiliation{Department of Physics, University of Surrey, Guildford, GU2 7XH, UK} 
\author{P. Navr\'{a}til}
\affiliation{TRIUMF, 4004 Wesbrook Mall, Vancouver, British Columbia, V6T 2A3, Canada}
\date{\today}         

\begin{abstract}
We derive \emph{ab initio} optical potentials from self-consistent Green's function (SCGF) theory and compute the elastic scattering of neutrons off oxygen and calcium isotopes.
The comparison with scattering data is satisfactory at low scattering energies.
The method is benchmarked against no-core shell model with continuum (NCSMC) calculations, showing that virtual excitations of the target are crucial to predict proper fragmentation and absorption at higher energies. 
This is a significant step toward deriving optical potentials for medium mass nuclei and complex many-body systems in general.
%%%enabling new insight into the scattering of complex many--body systems in general.
%The comparison with available elastic scattering data shows that it is possible to reproduce low energy scattering observables satisfactorily.
\end{abstract}
%\PACS{21.60.De, 24.10.Ht, 25.40.Dn}
\maketitle

\emph{Introduction.}
Reactions are
a fundamental aspect of nuclear physics since they are
used experimentally to determine many properties of atomic nuclei. They are also a key component to significant scientific questions, such as the reaction networks that control nucleosynthesis.  Unfortunately, first principles theoretical descriptions for scattering on medium-mass nuclei are still lacking. Even tough ground state properties and excited states can be calculated \emph{ab initio}, the complexity of many-body dynamics forces us to model the reaction mechanisms in terms of phenomenological optical potentials. This lack of consistency among the structure and reaction theories has been a major issue for nuclear physics for decades.

Optical models are an effective way to decouple the scattering wave function of the projectile from the internal structure of the target. Thus, microscopic (non-phenomenological) formalisms have also been proposed to compute them~\cite{Feshbach:58,Capuzzi:96}, although working implementations are still scarce. In this Letter, we discuss an \emph{ab initio} 
calculation of optical potentials that starts from saturating nuclear forces and compares favourably with low-energy scattering data. In doing so, we also identify key ingredients needed to improve the predictability at higher energies.
This represent a  successful step toward gaining insight into the reaction dynamics and to perform reliable predictions of scattering with exotic nuclei.

%Recent years have seen considerable advances in the theory of optical potentials. Non locality effects have been shown to be crucial for describing three-body processes~\cite{Timofeyuk:13,Bailey:16}, the importance of both scattering and bound states in the coupling to breakup channels has been  explored~\cite{Garcia:15}, and global dispersive optical potentials have been developed~\cite{Dickhoff:16}. 

%However, the greatest challenge remains the one of describing the nuclear structure and scattering consistently, from the same theory.
Many-body Green's function methods are particularly suited to pursue this goal for medium and heavy nuclei since their central quantity, the self-energy, is naturally linked to the Feshbach theory of optical potentials~\cite{Feshbach:58,Cederbaum:2001}. While the particle part of the self-energy is equivalent to the original formulation of Feshbach, its hole part also describes the structure of the target~\cite{Capuzzi:96}. Hence, it facilitates a consistent treatment of scattering and structure. %, based on the same microscopic Hamiltonian.

Some related (semi--)phenomenological attempts to exploit Green's function methods  include the nuclear field theory \cite{Mahaux:85,Idini:12} and its extension to nuclear transfer reactions \cite{Idini:15,Broglia:16}. Another incarnation of Green's function related theories is the dispersive optical model~\cite{Johnson:88}, which is a data driven formulation of global  (local and non local) potentials constructed as the best possible parameterization of a microscopic self-energy~\cite{Charity:06,Dickhoff:16}. The nuclear structure method was applied recently obtaining good reproduction of $^{40}$Ca scattering based on the Gogny D1S interaction~\cite{Blanchon:15}.
Other approaches, based on the nucleon-nucleon T-matrix and folding with the nuclear density have proven to be effective~\cite{Vorabbi2016PRC,Gennari2018PRC,Burrows:2018ggt,Whitehead:2018bfs}.

%
% It is an important comment but we have little space [CB, 5.12.2018]
%
%For the study of transfer reactions, such as $(d,p)$,  it would be particularly important to have an optical potential that is deduced consistently from the same Hamiltonian used in the proton-neutron channel~\cite{Timofeyuk:13,Garcia:15,Bailey:16}. To do so, one  needs  {\em realistic} nuclear interactions and {\em ab initio} calculations of elastic nucleon-nucleus scattering.
%

\emph{Ab initio} methods have been successful in direct calculations of scattering when only few nucleons are at play.
Quantum Monte Carlo has been historically used for light nuclei~\cite{Varga:02,Nollett:07,Lynn:16}. %} and was recently applied to alpha--nucleon scattering exploiting chiral three-nucleon (3N) forces~\cite{Lynn:16}.
The no-core shell model with resonating group method~(NCSM/RGM) or with \hbox{continuum~(NCSMC)} have been successful in calculating scattering and transfer reactions for light targets~\cite{Navratil:10, Baroni:13prl, Raimondi:16}. Coupled cluster theory has also been employed with a Gamow basis for \hbox{proton--$^{40}$Ca}~\cite{Hagen:12} and  combined with a Green's function approach to compute phase shifts for $^{16}$O and Ca isotopes~\cite{Rotureau:17,Rotureau:18}. 
On the other hand, the self-consistent Green's function (SCGF) formalism~\cite{Dickhoff:04,Soma:11} can \hbox{calculate} the microscopic optical potential directly even for heavier nuclei. This approach has been used to compute phase shifts~\cite{Barbieri:05} and to investigate analytical properties of optical models~\cite{Waldecker:11}. However, these early studies were limited to two-nucleon (NN) forces and a comparison to the experiment has been hindered by the lack of realistic Hamiltonians capable to reproduce the radius of the target.

Three-nucleon (3N) interactions have been recently formulated and implemented for SCGF theory in~\cite{Carbone:13,Cipollone:15,Raimondi20183nf}. Moreover, the introduction of saturating nuclear interactions~\cite{Ekstrom:15} has allowed a good reproduction of radii and binding energies across the oxygen~\cite{Lapoux:16} and calcium chains~\cite{Garcia:16}. Hence, we are now in the position to meaningfully compare first principle approaches to scattering data in medium-mass nuclei. 
In the following, we present state of the art SCGF calculations to test current {\em ab initio} methods and compare our results to NCSM/RGM and NCSMC computations with NN  and NN+3N interactions. 
We then use a saturating chiral Hamiltonian to study elastic scattering of neutrons from $^{16}$O and $^{40}$Ca.

%\section{The microscopic optical potential}

%\subsection{Subsection}
%The irreducible self-energy, $\Sigma^\star(\omega)$, has the following general spectral representation,
%  considering the propagators as solutions of the Dyson equation
% \begin{equation}
%  g_{\alpha,\beta} (E) = g^0_{\alpha,\beta} + \sum_{\gamma,\delta} g^0_{\alpha,\gamma} \Sigma^\star_{\gamma,\delta}(E) g_{\delta,\beta} (E)
% \end{equation}
%as,
\emph{Formalism.} The Hamiltonian used to compute the self-energy is
\begin{equation}
 H(A) = \hat{T} - \hat{T}_{c.m.}(A+1) + \hat{V} + \hat{W}
 \label{eq:Hint}
\end{equation}
where $\hat{T}_{c.m.}(A+1)$ is the center of mass kinetic energy for the $A$-nucleon target plus the projectile and $\hat{V}$ and $\hat{W}$ are the 
%two-- (NN) and three--body (3N) 
NN and 3N interactions.  $\hat{W}$ is included as an equivalent effective two--body interaction, averaged on the correlated propagator as discussed in Refs.~\cite{Cipollone:15,Rocco2018escatt}. 
The SCGF calculation proceeds by solving the Dyson equation, $g(\omega)=g^0 (\omega) + g^0(\omega) \Sigma^\star(\omega) g(\omega)$, in an harmonic oscillator (HO) basis of $N_{\rm max}$+$1$ shells, where $g^0(\omega)$ is the free particle propagator and the irreducible self-energy $\Sigma^\star(\omega)$ has the following general spectral representation:
\begin{align}
\Sigma^\star_{\alpha \beta}(E,&\Gamma) = \Sigma^{(\infty)}_{\alpha \beta} \! + \!  \sum_{i, j}  {\bf M}^{\dagger}_{\alpha, i} \!\! \left[ \frac{1}{E - (\textbf{K}^> \!\! + \textbf{C}) + i \Gamma} \right]_{i, j} \!\!\!{\bf M}_{j,\beta}
 \nonumber \\ 
 & + \sum_{r, s} {\bf N}_{\alpha, r} \! \left[ \frac{1}{E - (\textbf{K}^< \!\!+ \textbf{D}) - i \Gamma} \right]_{r, s} \!\!\! {\bf N}^{\dagger}_{s,\beta} \, ,
 \label{eq:Sigma_ho}
\end{align}
%In Eq.~\eqref{eq:Sigma_ho}, 
where $\alpha$ and $\beta$ label the single particle quantum numbers of the HO basis, $\Sigma^{(\infty)}$ is the correlated and energy independent mean field, and $\Gamma$ sets the correct boundary conditions.~% self-energy
We performed calculations with the third order algebraic diagrammatic construction [ADC(3)] method, where the matrices  $\bf{M}$~($\bf{N}$)  couple single particle states to intermediate 2p1h~(2h1p) configurations, ${\bf C}$~(${\bf D}$) are interaction matrices among these configurations, and $\bf{K}$ are their unperturbed energies \cite{Schirmer:83,Barbieri:17lnp}.
All intermediate 2p1h and 2h1p states (respectively labelled by indices $i,j$ and $r,s$) were included. For $N_{\rm max}=13$, this incorporates configurations up to  $400$ MeV of excitation energy and partial waves of the projectile up to angular momentum $j=27/2$ for both parities.

The resulting dressed single particle propagator can be written in the K\"all\'en–Lehmann representation as
\begin{align}
g_{\alpha \beta}(E,\Gamma) = & \sum_n \frac{\bra{\Psi^A_{0}} c_\alpha \ket{\Psi^{A+1}_n} \bra{\Psi^{A+1}_n} c^\dagger_{\beta} \ket{\Psi^A_{0}}}{E - E^{A+1}_n + E^{A}_0 + i\Gamma} \notag \\ 
  + & \sum_k \frac{\bra{\psi^A_{0}} c^\dagger_\alpha \ket{\Psi^{A-1}_k} \bra{\Psi^{A-1}_k} c_{\beta} \ket{\Psi^A_{0}}}{E - E^{A}_0 + E^{A-1}_k - i\Gamma} \; .
\end{align}
The poles of the forward-in-time propagator, $E^{A+1}_n-E^A_0$, indicate then the energy of the $n$th exited state of the $(A+1)$-nucleon system with respect to the ground state of the target $A$. Hence, they are directly identified with the scattering energy. For each many-body state $\ket{\Psi^{A+1}_n}$ in the continuum, the corresponding overlaps \hbox{$\psi_n(\alpha)\equiv\bra{\Psi^{A+1}_n} c^\dagger_\alpha \ket{\Psi^A_{0}}$} are associated with the elastic scattering wave function through Feshbach theory~\cite{Feshbach:58,Escher:02}.

Although the scattering waves are unbound, the self-energy $\Sigma^{\star}(\omega)$ associated with the optical potential is localized and it can be efficiently expanded on square integrable functions. 
 Hence, we proceed by calculating Eq.~\eqref{eq:Sigma_ho} in HO basis but transform it to momentum space before solving the scattering problem. This will ensure that the proper asymptotic behaviours of both bound and scattering states are obtained.
The optical potential for a given partial wave ($l,j$) is then expressed as
\begin{equation}
\Sigma^{\star  \,  l,j}(k, k';  E, \Gamma) = \sum_{n,n'} R_{n,l}(k)  \Sigma^{\star \, l,j}_{n,n'}(E, \Gamma)  R_{n',l}(k') \, , 
\label{eq:sigma}
\end{equation}
which is non local and energy-dependent and where $R_{n,l}(k)$ are the radial HO wavefunctions in momentum space.
Through Eqs.~\eqref{eq:Sigma_ho} and~\eqref{eq:sigma}, the SCGF approach provides a parametrized, separable and analytical form of the optical potential. % within the Lehmann representation.

The parameter $\Gamma$ sets the time ordering boundary conditions, but it does not affect the solution of the many-body problem that comes from the diagonalization of the equation of motion \cite{Barbieri:05,Idini:12,Barbieri:17lnp}.  However, we retain it in Eq.~\eqref{eq:sigma} to introduce a small finite width for the 2p1h/2h1p configurations, which would otherwise be discretised in the present approach. We checked that this does not affect our conclusions below.
% In the scattering calculations below, we use $\Gamma(E)=0.002 \textrm{MeV} \epsilon^2/\pi((E-E_F)^2-(22.36 \textrm{MeV})^2)$, where $E_F$ is the Fermi energy, and find that this does not affect the scattering results discussed below.

We use  the intrinsic Hamiltonian of Eq.~\eqref{eq:Hint} and large enough HO spaces so that the intrinsic ground state decouples from the center of mass motion~\cite{Hagen:09}. Even if decoupled, the latter is not fully suppressed and the self-energy~\eqref{eq:sigma} is still computed in laboratory frame. We correct for this by rescaling the scattering momentum appropriately, which naturally leads to the correct center of mass energy $E_{c.m.}$ and reduced mass $\mu = \gamma m$, with $\gamma \equiv A/(A+1)$. The Dyson equation eventually reduces to the following one-body eigenvalue problem~\cite{Dickhoff:04,Barbieri:17lnp}:
\begin{align}
 \left[ E_{c.m.} - k^2/(2\mu) \right] &\; \psi_{l,j} (k) =  \nonumber \\
 \int\textrm{d}k' k'^2  \gamma^3  &\; \Sigma^{\star  \,  l,j}\left(\gamma k, \gamma k'; \gamma E_{c.m.}, \Gamma \right) \psi_{l,j} (k'),
 \label{eq:Schroedinger}
\end{align}
We diagonalize this Schr\"odinger--like equation in momentum space so that the kinetic energy is treated exactly 
%--without truncations in HO space--%
and we account for the non locality and $l,j$ dependence of Eq.~\eqref{eq:sigma}. 
The phase shifts $\delta(E_{c.m.})$ are obtained as function of the projectile energy, for each partial wave, from where the differential cross section can be calculated.
The bound states solutions of Eq.~\eqref{eq:Schroedinger} yields overlap wave functions between $\ket{\Psi^A}$ and $\ket{\Psi^{A+1}}$~\cite{Dieperink:74}. Hence, they provide spectroscopic factors and asymptotic normalization coefficients that can be employed for the consistent computation of nucleon capture and knockout processes.

\begin{figure}[t]
%\centerline{%
\includegraphics[width=0.45\textwidth]{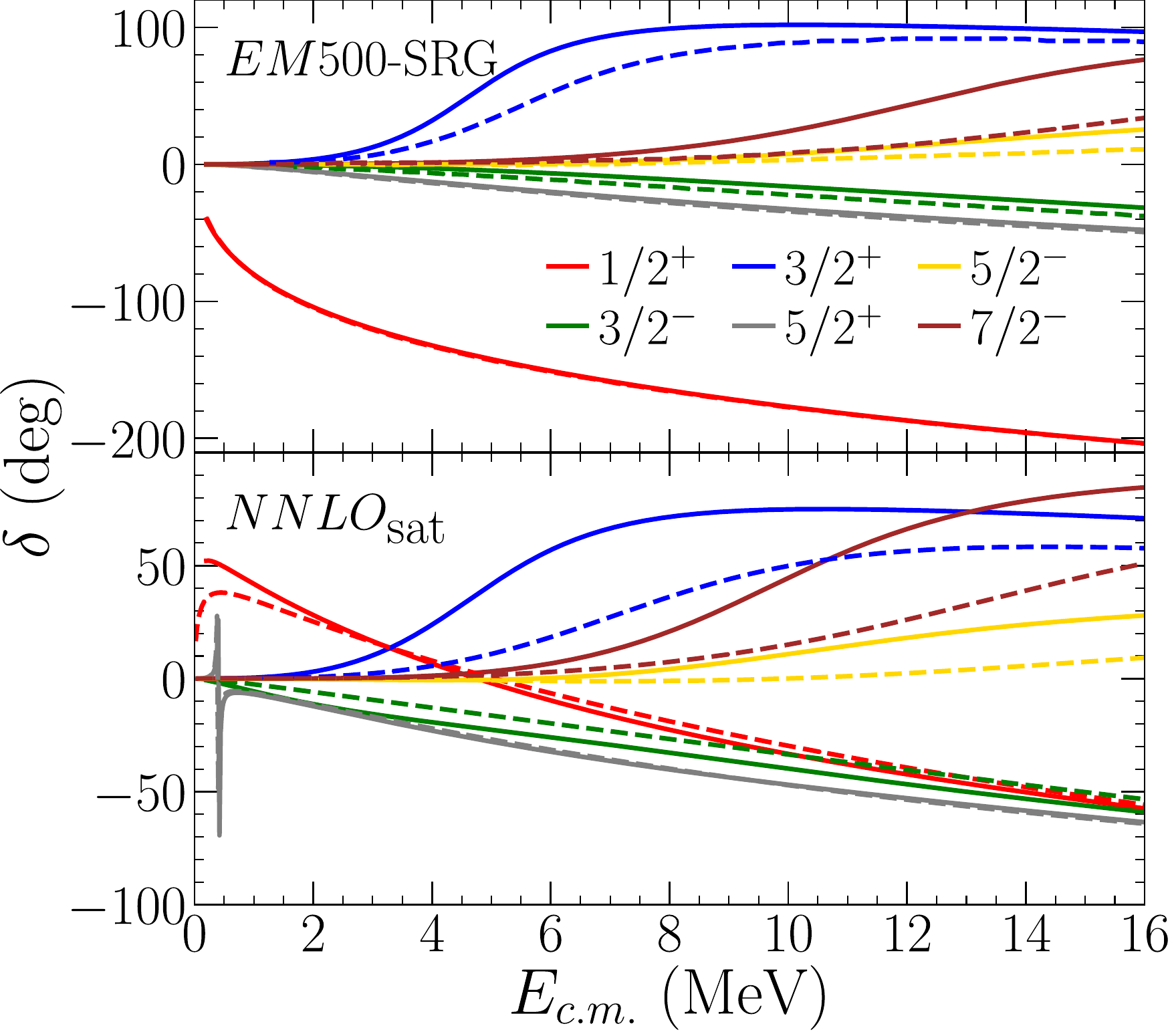} % \hspace{0.02\textwidth}
%} %450 words
\caption{Real part of nuclear phase shifts, $\delta(E_{c.m.})$, for neutrons scattering off $^{16}$O as a function of energy
obtained from the EM500-SRG (upper panel) and the NNLO$_{\textrm{sat}}$ (lower panel) interactions.
The solid lines are SCGF calculations using only the static part of the self-energy $\Sigma^{(\infty)}$ in a $N_{\rm max}=13$ space. Dashed lines are for NCSM/RGM, which included only the ground state of $^{16}$O and used a no-core model space up to $N_{\rm NCSM} = 18\,\hbar\Omega$ (top, form Ref.~\cite{Navratil:10}) and $8\,\hbar\Omega$ (bottom). % as in \cite{Navratil:10}.
}
\label{Fig:comparison}
\end{figure}
\emph{Results.} 
We first compare to early NCSM/RGM results from Ref.~\cite{Navratil:10}, where neutron scattering off $^{16}$O was computed with a NN-only interaction derived from the chiral N$^3$LO force of Ref.~\cite{Entem:03} (EM500) and evolved with free space similarity renormalization group (SRG)~\cite{Bogner:07} to a cutoff~$\lambda$~=~2.66~fm$^{-1}$.  This soft interaction facilitates model space convergence and allows for a 
more meaningful benchmark.
These early NCSM/RGM computations did not include virtual excitations of the target nucleus. For consistence, we performed our SCGF calculations with the same Hamiltonian but evaluated  the phase shifts using only the static self-energy, $\Sigma^{(\infty)}$. 
The comparison is shown by the upper panel of Fig.~\ref{Fig:comparison} and it is very satisfactory for the $j^\pi=1/2^+$ and $5/2^+$ partial waves. For this light nucleus, the discrepancy of about 1~MeV for the energy of the $3/2^+$ resonance is also consistent with the uncertainty in the transformation to the center of mass system done in Eq.~\eqref{eq:Schroedinger}.
As we discuss below, doorway excitations of the target nucleus have a strong impact on the energies of single particle resonances. 
To account for this, we performed new NCSMC calculations that can also include low-lying excitations of $^{17}$O. Extrapolating from model spaces of $N_{\rm NCSM}$ = 6--10\,$\hbar\Omega$ we find quasiparticle energies of -3.4, -2.7, and 3.2~MeV for the $5/2^+, 1/2^+$ bound states and the $3/2^+$ resonance, respectively.  The corresponding results from SCGF, including the full $\Sigma^{\star}(\omega)$ self-energy, are $-6.3$, $-5.5$, and $0.5$ MeV. 
These should be expected to be more bound since SCGF introduces a larger number of 2p1h doorway configurations. At the same, time the excitation energies relative to the $^{17}$O ground state agree to within 200~keV, which is a satisfactory agreement given the different many-body truncations of the two approaches.

%The analogous comparison for the chiral NNLO$_{\textrm{sat}}$ NN+3N interaction of Ref.~\cite{Ekstrom:15} is harder to converge. Even with importance-truncation techniques~\cite{Roth2007it} we limited our NCSM/RGM calculations to $N_{\rm NCSM}$ = 8\,$\hbar\Omega$, estimating an uncertainty of 1--2~MeV for the position of esonances. % due to the model space truncation. 
We performed an analogous comparison for the chiral NNLO$_{\textrm{sat}}$ NN+3N interaction of Ref.~\cite{Ekstrom:15}.
For NCSM techniques, $^{16}$O is more difficult to converge because the interaction is harder and the additional 3N matrix elements limit the applicability of importance-truncation~\cite{Roth2007it}. We performed our NCSM/RGM calculations at $N_{\rm NCSM}$\,=\,8\,$\hbar\Omega$, and estimated an uncertainty of 1--2~MeV for the position of resonances. % due to the model space truncation. 
The SCGF still allows computations with $N_{\rm max}=13$ and we find that phase shifts are well converged up to 15~MeV for this space. This puts in evidence the advantage of the latter approach to address \emph{ab initio} scattering off medium mass isotopes.
The NNLO$_{\textrm{sat}}$ benchmark is displayed by the lower panel of Fig.~\ref{Fig:comparison} and it is qualitatively similar to the case of the soft EM500-SRG interaction, with the $j^\pi=1/2^+$ and $5/2^+$ waves agreeing best. For both Hamiltonians, the largest discrepancies are for the $j^\pi=3/2^+$ and~$7/2^-$ resonances, which are more affected by correlations in the continuum and the different many-body truncations of the  two approaches.
 NNLO$_{\textrm{sat}}$ was explicitly constructed to reproduce correct nuclear saturation properties of medium mass nuclei, including binding energies and radii.  The constraint on radii is crucial to predict elastic scattering observables that can be reasonably compared to the experiment, hence we will focus on this Hamiltonian in the following.

%
%\begin{figure}[t]
%\centerline{
%%\includegraphics[width=0.50\textwidth]{Figs/PhSh_O16_vs1} % \hspace{0.02\textwidth}
%%\includegraphics[width=0.25\textwidth]{Figs/PhSh_O16_vd3_MF} \includegraphics[width=0.25\textwidth]{Figs/PhSh_O16_vs1_MF} 
%\includegraphics[width=0.24\textwidth]{Figs/PS_conv_s1}  \includegraphics[width=0.24\textwidth]{Figs/PS_conv_d3}
%} 
%\caption{(Color online) Phase shifts, $\delta(E_{c.m.})$, for neutron scattering off $^{16}$O in the $s_{1/2}$ (left panel) and $d_{3/2}$ (right panel) partial waves. The sole static self-energy, $\Sigma^{(\infty)}$, was used with the NNLO$_\textrm{sat}$ interaction. Different model spaces with frequency $\hbar\Omega$=~20~MeV and sizes $N_{\rm max}=$~7 to 13 are shown.
%%{\em Left panel}: dependence of the  $s_{1/2}$ partial wave on the number of oscillator shells, for $N_{\rm max}$=11 (dashed) and 13 (solid). The oscillations at larger energies are narrow resonances. 
%%The dotted line has been calculated with the full ADC(3) $N_{\rm max}=13$ calculation, and then building the optical potential with an additional cutoff $N_{\rm cut}=11$.
%}
%\label{Fig:phsh}
%\end{figure}

\begin{figure}[t!]
%\centerline{%
\includegraphics[width=0.48\textwidth]{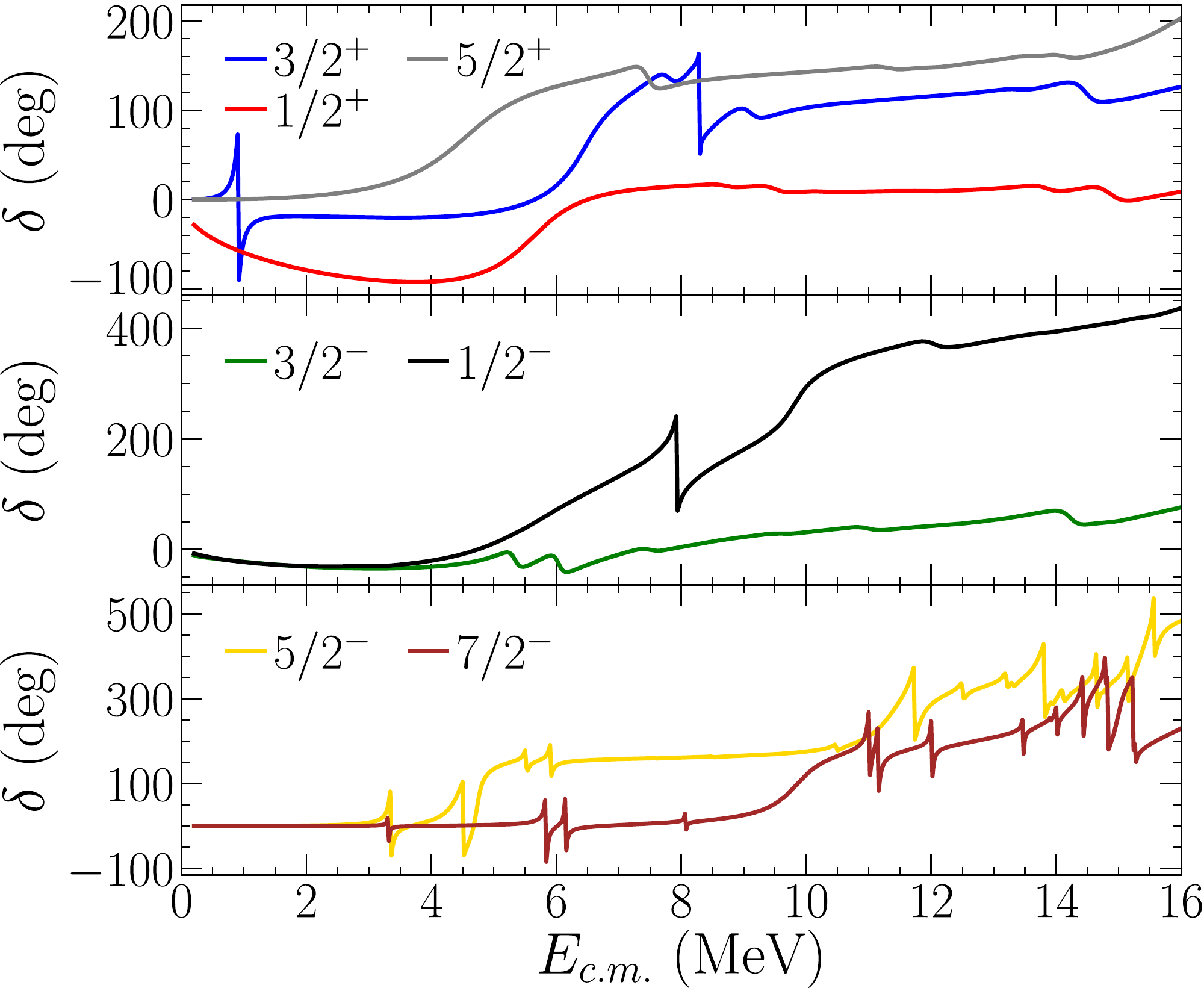}
%\includegraphics[width=0.4\textwidth]{Figs/PhaseShift_SE_+}
%\includegraphics[width=0.4\textwidth]{Figs/PhaseShift_SE_p}
%\includegraphics[width=0.4\textwidth]{Figs/PhaseShift_SE_f} 
%}
\caption{Real phase shifts, $\delta(E_{c.m.})$, for neutrons scattering off $^{16}$O using the complete self-energy,  Eq.~\eqref{eq:Sigma_ho}, and NNLO$_\textrm{sat}$ in an oscillator space of frequency $\hbar\Omega$~=~20~MeV and size $N_{\rm max}$~=~13. Positive parity   (upper panel), $l$=1  (central panel) and $l$=3 partial waves (lower panel) are shown.
}
\label{Fig:comparison_SE}
\end{figure}

Virtual excitations of the target have the double effect of increasing the attraction of the real part of the optical potential (hence, lowering the single particle spectrum) and of generating a large number of narrow resonances. This is clearly seen in Figs.~\ref{Fig:comparison_SE} that displays the phase shifts for neutron elastic scattering predicted by the whole self-energy of Eq.~\eqref{eq:Sigma_ho}. Most of the virtual excitations responsible for this, especially at low energy, are accessed by coupling to hundreds of 2p1h configurations for $^{17}$O and appear as clear spikes or ``smoothed'' oscillations in the figure. The SCGF-ADC(3) approach has the advantage of including these states naturally, even to large energies, so it describes efficiently the relevant physics.
Table~\ref{Table:1} compares the energies of some representative bound and scattering states to the experiment.
The $3/2^+$ single particle resonance is computed at \hbox{$0.91$~MeV} in the c.o.m. frame, very close the experimental value.
The first $1/2^-$ and $3/2^-$ are both predicted as bound states, although experimentally they are found inverted with the $3/2^-$ in the continuum. 
%
%We note that the NCSMC predicts the correct ordering for these two states but places them at much higher energies than the observation. 
%
We calculate a narrow width for a $5/2^-$ and a $7/2^-$ resonances, corresponding to excited states, close to the ones observed at 3.02 and 3.54 MeV~\cite{Lister:66}. However, there are other very narrow  {$f$-wave} resonances, measured between \hbox{1.55-2.82~MeV}, that our SCGF calculations do not resolve.
In general, we find that NNLO$_{\rm sat}$  predicts the location of dominant quasiparticle and holes states with an accuracy of~$\lesssim$~1~MeV for this nucleus.

\begin{table} [b]
\begin{center}
\begin{tabular}{lccccccccc}
\!\!\!\!$\varepsilon$ (MeV)
                                                     & $5/2^+$ & $1/2^+$ & $1/2^-$ & $5/2^-$ & $3/2^-$ & $3/2^+$ & $5/2^+_*$ & $5/2^-_*$ & $7/2^-_*$ \\
%$\varepsilon$ (MeV)          & $d_{5/2}$ & $s_{1/2}$ & $p_{1/2}$ & $f_{5/2}$& $d_{3/2}$ & $p_{3/2}$ & $d_{5/2}^*$ & $f_{5/2}^*$ & $f_{7/2}^*$ \\
\hline
\!\!\!\!exp.                                    & -4.14   & -3.27   & -1.09 &  -0.30   &  0.41 &  0.94    & 3.23    & 3.02    & 3.54    \\
\!\!\!\!NNLO$_{\rm sat} \!\!\!$          
                                           & -5.06   & -3.58   & -0.15   & -1.23  & -2.24 &  0.91    & 4.57    & 3.36    & 3.37    \\
\hline
\end{tabular}
\caption{
Excitation spectrum of $^{17}$O with respect to the $n$+$^{16}$O threshold, as obtained from Eq. (\ref{eq:Schroedinger}) and the NNLO$_{\rm sat}$ interaction and compared to the experiment~\cite{Tilley:93}. Broad resonances in the continuum (most notably, the $5/2^+$) are computed at midpoint. 
The asterisks $(_*)$ indicate higher excited states, above the lowest one, for each partial wave.}
\label{Table:1}
\end{center}
\end{table}
\begin{figure}[t]
\centerline{
\includegraphics[width=0.48\textwidth]{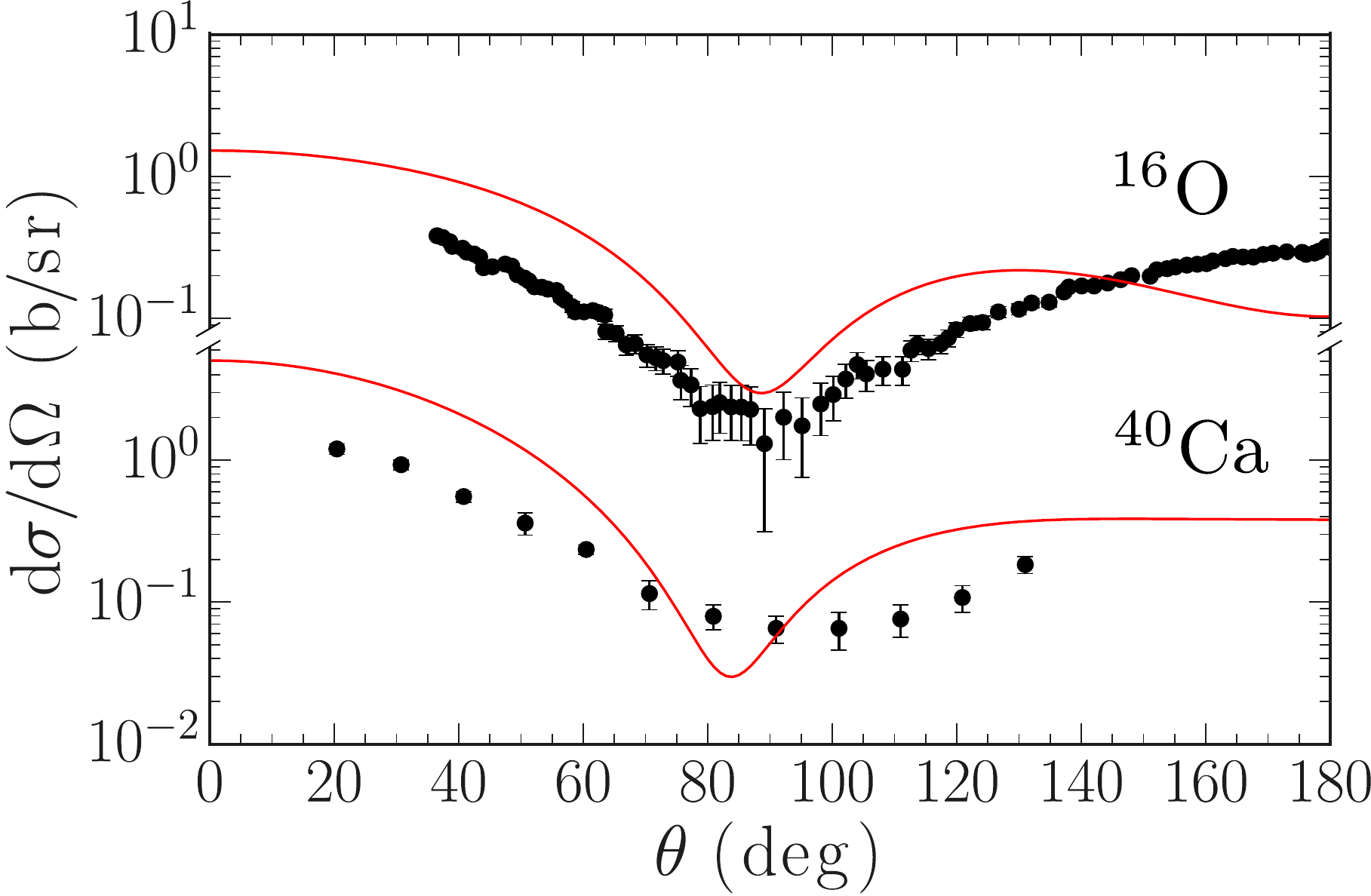}}
\caption{Differential  cross section for neutron elastic scattering off $^{16}$O ( $^{40}$Ca) at 3.286 (3.2) MeV of neutron energy, with NNLO$_\textrm{sat}$ and compared to the empirical data from~\cite{Lister:66,Becker:66}.
} 
\label{Fig:dSig_nO16_nCa40}
\end{figure}

\begin{figure}
\centerline{
\includegraphics[width=0.48\textwidth]{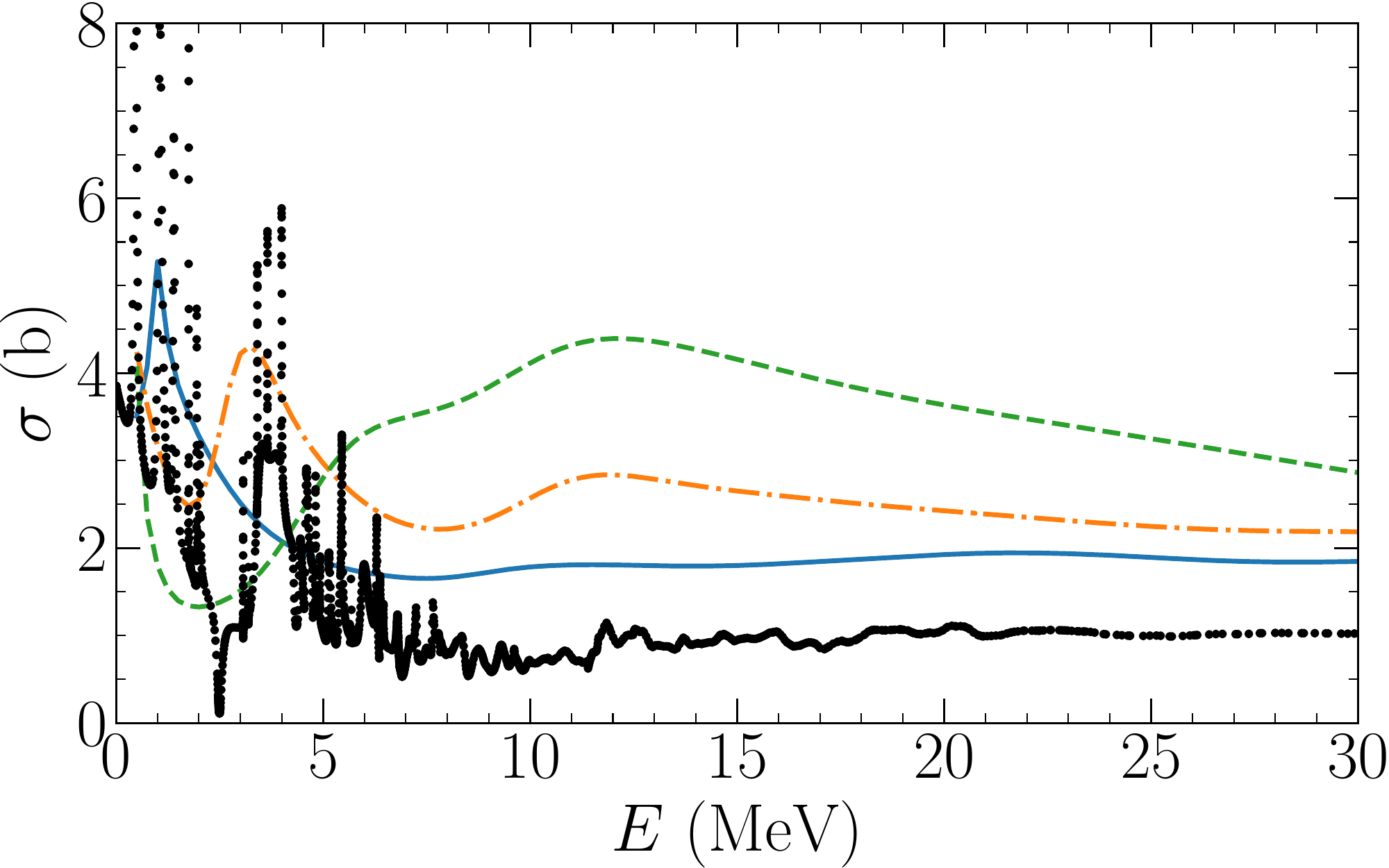}}
\caption{Total elastic cross section for neutron elastic scattering on $^{16}$O form SCGF-ADC(3) at different incident neutron energies, compared to the experiment from~\cite{ENDF2018}. The dashed, dot-dashed and full lines correspond to the sole static self-energy $\Sigma^{(\infty)}$, to  retaining 50\% of the 2p1h/2h1p doorway configurations and to the complete Eq.~\eqref{eq:Sigma_ho}, respectively.
} 
\label{Fig:Sig_tot}
\end{figure}

Fig.~\ref{Fig:dSig_nO16_nCa40} compares the low-energy differential cross sections originating from Eq.~\eqref{eq:Schroedinger} to neutron scattering data for $^{16}$O at 3.286~MeV and $^{40}$Ca at 3.2~MeV.
The minima are reproduced well for $^{16}$O (and close to the experiment for $^{40}$Ca), confirming the correct prediction of density distributions for NNLO$_{\rm sat}$~\cite{Ekstrom:15,Garcia:16,Lecluse:17}.
However, results are somewhat overestimated and hint at a general lack of absorption that is usually faced by attempts at computing the optical potentials from~\emph{ab initio}. This is likely related to missing doorway configurations (3p2h and beyond) that should be propagated in the denominators of Eq.~\eqref{eq:Sigma_ho} but are neglected by state of the art approaches.  Note that there are more than 200 experimentally observed excitations already between the ground state and the neutron separation threshold in $^{41}$Ca \cite{Nesaraja:16}, while the SCGF-ADC(3) predicts only about 40 of them. This issue is likely to worsen at higher energies where configurations more complex than 2p1h become relevant.
We investigated this problem by computing total $n$+$^{16}$O elastic cross sections, $\sigma(E_{c.m.})$, with only $\Sigma^{(\infty)}$,  suppressing 50\% of 2p1h/2h1p states (evenly across all energies), and by using the complete ADC(3) self-energy. Fig.~\ref{Fig:Sig_tot} shows that $\sigma(E_{c.m.})$ presents oscillations up to about 5~MeV. These are in part reproduced by theory and are sensible to interferences among the projectile and the included 2p1h configurations. However, the link between absorption and the density of intermediate doorway configurations becomes clear at higher energies and it is confirmed by our calculations~\footnote{Note that the Lanczos algorithm used the solve for Eq.~(\ref{eq:Sigma_ho}) does not affect these conclusions since it is specifically designed to preserve the strength distribution of response functions~\cite{Waldecker:11,Soma:14}.}.

%\begin{figure}
%\centerline{
%\includegraphics[width=0.48\textwidth,height=0.30\textwidth]{Figs/O16_CS_3-286MeV}}
%\caption{(Color online) Differential  cross section for neutron elastic scattering on $^{16}$O at 3.286 MeV of neutron energy, with NNLO$_\textrm{sat}$ and compared with experimental data from~\cite{Lister:66}. 
%}
%\label{Fig:dSig_nO16}
%\end{figure}
%%
%%
%\begin{figure}
%\centerline{
%\includegraphics[width=0.45\textwidth,height=0.30\textwidth]{Figs/Ca40_CS_3-2MeV}}
%\caption{(Color online) Differential  cross section for neutron elastic scattering on $^{40}$Ca at 3.2 MeV of neutron energy, with NNLO$_\textrm{sat}$ and compared with experimental data from~\cite{Becker:66}.
%} 
%\label{Fig:dSig_nCa40}
%\end{figure}

%\emph{Conclusions.}
To conclude, we have benchmarked optical potentials generated through SCGF theory to analogous full scale NCSMC simulations and to data for neutron elastic scattering at low energy.  For both theory approaches, the correct asymptotic behaviour of the scattering wave are reproduced even if the target wave function and the optical potentials are expanded in a HO basis.  The theory benchmark, with freezing of virtual excitation of the target, is very encouraging. The SCGF approach also has the capability of accounting for a large number of such intermediate excitations up to very large energies, and it achieves a promising description of complex resonance states from first principles.
The use of a saturating chiral interaction allows us to make a meaningful comparison to the experiment, which was not possible in previous investigation of this approach.
Overall, we found that the most important features of optical potentials at low energy are well reproduced, together with key observables related to elastic scattering.

The present study also puts in evidence how the lack of absorption normally observed in \emph{ab initio} generated optical potentials is directly linked to the neglect of doorway configurations beyond 2p1h ones. Thus, addressing this challenge will be the next fundamental step toward predictive theories at medium scattering energies.  
It remains clear from the present results that obtaining reliable \hbox{\em ab initio} of optical potentials, directly from the self-energy, is becoming a goal within reach. The present findings open a path to establish consistent theories of structure and reactions for medium--mass nuclei.

%This will enable us to study nuclear reactions in a novel way, building nuclei from the interactions based on first principles chiral symmetries, up to direct reactions in finite nuclei, which has the considerable advantage of not being biased by data in the mass region under consideration. Thus the reliability of such calculation is similar for well known $^{40}$Ca as for exotic neutron rich systems. %Calculations for Oxygen and Calcium isotopes at subshell closure have already been performed.
%Extensive results for targets along the subshell closure of O and Ca chains will be presented in a forthcoming publication.

%\section*{Acknowledgments}
~\\
\emph{Acknowledgments}
A.I. was supported by the Royal Society and Newton Fund through the Newton International Fellowship No. NF150402.
This work was supported the United Kingdom Science and Technology Facilities Council (STFC) under Grants No. ST/P005314/1 and No.
ST/L005816/1 and by the NSERC Grant No. SAPIN-2016-00033. 
TRIUMF receives federal funding via a contribution agreement with the National Research Council of Canada.
Computations were performed using the DiRAC Data Intensive service at Leicester (funded by the UK BEIS via STFC capital grants ST/K000373/1 and ST/R002363/1 and STFC DiRAC Operations grant ST/R001014/1)  and an INCITE Award on the Titan supercomputer of the Oak Ridge Leadership Computing Facility (OLCF) at ORNL.

%\clearpage
%\section*{References}
\bibliographystyle{iopart-num}

\bibliography{./nuclear}

\providecommand{\newblock}{}
\begin{thebibliography}{10}
\expandafter\ifx\csname url\endcsname\relax
  \def\url#1{{\tt #1}}\fi
\expandafter\ifx\csname urlprefix\endcsname\relax\def\urlprefix{URL }\fi
\providecommand{\eprint}[2][]{\url{#2}}

\bibitem{Feshbach:58}
Feshbach H 1958 {\em Annals of Physics\/} {\bf 5} 357 -- 390 ISSN 0003-4916
  \urlprefix\url{http://www.sciencedirect.com/science/article/pii/0003491658900071}

\bibitem{Capuzzi:96}
Capuzzi F and Mahaux C 1996 {\em Annals of Physics\/} {\bf 245} 147 -- 208

\bibitem{Cederbaum:2001}
Cederbaum L~S 2001 {\em Ann. Phys.\/} {\bf 291} 169 -- 201

\bibitem{Mahaux:85}
Mahaux C, Bortignon P~F, Broglia R~A and Dasso C~H 1985 {\em Phys. Rep.\/} {\bf
  120} 1

\bibitem{Idini:12}
Idini A, Barranco F and Vigezzi E 2012 {\em Phys. Rev. C\/} {\bf 85} 014331
  \urlprefix\url{http://link.aps.org/doi/10.1103/PhysRevC.85.014331}

\bibitem{Idini:15}
Idini A, Potel G, Barranco F, Vigezzi E and Broglia R~A 2015 {\em Phys. Rev.
  C\/} {\bf 92}(3) 031304
  \urlprefix\url{http://link.aps.org/doi/10.1103/PhysRevC.92.031304}

\bibitem{Broglia:16}
Broglia R~A, Bortignon P~F, Barranco F, Vigezzi E, Idini A and Potel G 2016
  {\em Physica Scripta\/} {\bf 91} 063012
  \urlprefix\url{http://stacks.iop.org/1402-4896/91/i=6/a=063012}

\bibitem{Johnson:88}
Johnson C~H and Mahaux C 1988 {\em Phys. Rev. C\/} {\bf 38} 2589

\bibitem{Charity:06}
Charity R~J, Sobotka L~G and Dickhoff W~H 2006 {\em Phys. Rev. Lett.\/} {\bf
  97}(16) 162503
  \urlprefix\url{http://link.aps.org/doi/10.1103/PhysRevLett.97.162503}

\bibitem{Dickhoff:16}
Dickhoff W~H, Charity R~J and Mahzoon M~H 2017 {\em J. Phys. G\/} {\bf 44}
  033001 \urlprefix\url{http://stacks.iop.org/0954-3899/44/i=3/a=033001}

\bibitem{Blanchon:15}
Blanchon G, Dupuis M, Arellano H~F and Vinh~Mau N 2015 {\em Phys. Rev. C\/}
  {\bf 91}(1) 014612
  \urlprefix\url{http://link.aps.org/doi/10.1103/PhysRevC.91.014612}

\bibitem{Vorabbi2016PRC}
Vorabbi M, Finelli P and Giusti C 2016 {\em Phys. Rev. C\/} {\bf 93}(3) 034619
  \urlprefix\url{https://link.aps.org/doi/10.1103/PhysRevC.93.034619}

\bibitem{Gennari2018PRC}
Gennari M, Vorabbi M, Calci A and Navr\'atil P 2018 {\em Phys. Rev. C\/} {\bf
  97}(3) 034619
  \urlprefix\url{https://link.aps.org/doi/10.1103/PhysRevC.97.034619}

\bibitem{Burrows:2018ggt}
Burrows M, Elster C, Weppner S~P, Launey K~D, Maris P, Nogga A and Popa G 2018
  (\textit{Preprint} \eprint{1810.06442})

\bibitem{Whitehead:2018bfs}
Whitehead T~R, Lim Y and Holt J~W 2018  (\textit{Preprint} \eprint{1812.08725})

\bibitem{Varga:02}
Varga K, Pieper S~C, Suzuki Y and Wiringa R~B 2002 {\em Phys. Rev. C\/} {\bf
  66}(3) 034611
  \urlprefix\url{https://link.aps.org/doi/10.1103/PhysRevC.66.034611}

\bibitem{Nollett:07}
Nollett K~M, Pieper S~C, Wiringa R~B, Carlson J and Hale G~M 2007 {\em Phys.
  Rev. Lett.\/} {\bf 99}(2) 022502
  \urlprefix\url{https://link.aps.org/doi/10.1103/PhysRevLett.99.022502}

\bibitem{Lynn:16}
Lynn J~E, Tews I, Carlson J, Gandolfi S, Gezerlis A, Schmidt K~E and Schwenk A
  2016 {\em Phys. Rev. Lett.\/} {\bf 116}(6) 062501
  \urlprefix\url{https://link.aps.org/doi/10.1103/PhysRevLett.116.062501}

\bibitem{Navratil:10}
Navr\'atil P, Roth R and Quaglioni S 2010 {\em Phys. Rev. C\/} {\bf 82}(3)
  034609 \urlprefix\url{http://link.aps.org/doi/10.1103/PhysRevC.82.034609}

\bibitem{Baroni:13prl}
Baroni S, Navr\'atil P and Quaglioni S 2013 {\em Phys. Rev. Lett.\/} {\bf
  110}(2) 022505
  \urlprefix\url{http://link.aps.org/doi/10.1103/PhysRevLett.110.022505}

\bibitem{Raimondi:16}
Raimondi F, Hupin G, Navr\'atil P and Quaglioni S 2016 {\em Phys. Rev. C\/}
  {\bf 93}(5) 054606
  \urlprefix\url{http://link.aps.org/doi/10.1103/PhysRevC.93.054606}

\bibitem{Hagen:12}
Hagen G and Michel N 2012 {\em Phys. Rev. C\/} {\bf 86}(2) 021602
  \urlprefix\url{http://link.aps.org/doi/10.1103/PhysRevC.86.021602}

\bibitem{Rotureau:17}
Rotureau J, Danielewicz P, Hagen G, Nunes F~M and Papenbrock T 2017 {\em Phys.
  Rev. C\/} {\bf 95}(2) 024315
  \urlprefix\url{http://link.aps.org/doi/10.1103/PhysRevC.95.024315}

\bibitem{Rotureau:18}
Rotureau J, Danielewicz P, Hagen G, Jansen G~R and Nunes F~M 2018 {\em Phys.
  Rev. C\/} {\bf 98}(4) 044625
  \urlprefix\url{https://link.aps.org/doi/10.1103/PhysRevC.98.044625}

\bibitem{Dickhoff:04}
Dickhoff W and Barbieri C 2004 {\em Progress in Particle and Nuclear Physics\/}
  {\bf 52} 377 -- 496 ISSN 0146-6410
  \urlprefix\url{http://www.sciencedirect.com/science/article/pii/S0146641004000535}

\bibitem{Soma:11}
Som\`a V, Duguet T and Barbieri C 2011 {\em Phys. Rev. C\/} {\bf 84}(6) 064317
  \urlprefix\url{http://link.aps.org/doi/10.1103/PhysRevC.84.064317}

\bibitem{Barbieri:05}
Barbieri C and Jennings B~K 2005 {\em Phys. Rev. C\/} {\bf 72} 014613

\bibitem{Waldecker:11}
Waldecker S, Barbieri C and Dickhoff W~H 2011 {\em Phys. Rev. C\/} {\bf 84}
  034616

\bibitem{Carbone:13}
Carbone A, Cipollone A, Barbieri C, Rios A and Polls A 2013 {\em Phys. Rev.
  C\/} {\bf 88}(5) 054326
  \urlprefix\url{http://link.aps.org/doi/10.1103/PhysRevC.88.054326}

\bibitem{Cipollone:15}
Cipollone A, Barbieri C and Navr\'atil P 2015 {\em Phys. Rev. C\/} {\bf 92}(1)
  014306 \urlprefix\url{http://link.aps.org/doi/10.1103/PhysRevC.92.014306}

\bibitem{Raimondi20183nf}
Raimondi F and Barbieri C 2018 {\em Phys. Rev. C\/} {\bf 97}(5) 054308
  \urlprefix\url{https://link.aps.org/doi/10.1103/PhysRevC.97.054308}

\bibitem{Ekstrom:15}
Ekstr{\"o}m A, Jansen G, Wendt K, Hagen G, Papenbrock T, Carlsson B,
  Forss{\'e}n C, Hjorth-Jensen M, Navr{\'a}til P and Nazarewicz W 2015 {\em
  Phys. Rev. C\/} {\bf 91} 051301

\bibitem{Lapoux:16}
Lapoux V, Som\`a V, Barbieri C, Hergert H, Holt J~D and Stroberg S~R 2016 {\em
  Phys. Rev. Lett.\/} {\bf 117}(5) 052501
  \urlprefix\url{http://link.aps.org/doi/10.1103/PhysRevLett.117.052501}

\bibitem{Garcia:16}
Garcia~Ruiz R~F and {\em et al} 2016 {\em Nat Phys\/} {\bf 12} 594--598
  \urlprefix\url{http://dx.doi.org/10.1038/nphys3645}

\bibitem{Rocco2018escatt}
Rocco N and Barbieri C 2018 {\em Phys. Rev. C\/} {\bf 98}(2) 025501
  \urlprefix\url{https://link.aps.org/doi/10.1103/PhysRevC.98.025501}

\bibitem{Schirmer:83}
Schirmer J, Cederbaum L~S and Walter O 1983 {\em Phys. Rev. A\/} {\bf 28}(3)
  1237 \urlprefix\url{http://link.aps.org/doi/10.1103/PhysRevA.28.1237}

\bibitem{Barbieri:17lnp}
Barbieri C and Carbone A 2017 {\em {\normalfont in} An Advanced Course in
  Computational Nuclear Physics: Bridging the Scales from Quarks to Neutron
  Stars\/} edited by M. Hjorth-Jensen, M.P. Lombardo, and U. van Kolck, Lecture
  Notes in Physics Vol. 936 (Springer)
  \urlprefix\url{https://books.google.co.uk/books?id=1rPU1Zz5wbMC}

\bibitem{Escher:02}
Escher J and Jennings B~K 2002 {\em Phys. Rev. C\/} {\bf 66} 034313

\bibitem{Hagen:09}
Hagen G, Papenbrock T and Dean D~J 2009 {\em Phys. Rev. Lett.\/} {\bf 103}(6)
  062503
  \urlprefix\url{https://link.aps.org/doi/10.1103/PhysRevLett.103.062503}

\bibitem{Dieperink:74}
Dieperink A~E~L and Forest T~d 1974 {\em Phys. Rev. C\/} {\bf 10}(2) 543--549
  \urlprefix\url{https://link.aps.org/doi/10.1103/PhysRevC.10.543}

\bibitem{Entem:03}
Entem D~R and Machleidt R 2003 {\em Phys. Rev. C\/} {\bf 68}(4) 041001
  \urlprefix\url{http://link.aps.org/doi/10.1103/PhysRevC.68.041001}

\bibitem{Bogner:07}
Bogner S~K, Furnstahl R~J and Perry R~J 2007 {\em Phys. Rev. C\/} {\bf 75}(6)
  061001 \urlprefix\url{https://link.aps.org/doi/10.1103/PhysRevC.75.061001}

\bibitem{Roth2007it}
Roth R and Navr\'atil P 2007 {\em Phys. Rev. Lett.\/} {\bf 99}(9) 092501
  \urlprefix\url{https://link.aps.org/doi/10.1103/PhysRevLett.99.092501}

\bibitem{Lister:66}
Lister D and Sayres A 1966 {\em Phys. Rev.\/} {\bf 143} 745

\bibitem{Tilley:93}
Tilley D, Weller H and Cheves C 1993 {\em Nuclear Physics A\/} {\bf 564} 1 --
  183 ISSN 0375-9474
  \urlprefix\url{http://www.sciencedirect.com/science/article/pii/0375947493900737}

\bibitem{Becker:66}
Becker R, Guindon W and Smith G 1966 {\em Nuclear Physics\/} {\bf 89} 154 ISSN
  0029-5582
  \urlprefix\url{http://www.sciencedirect.com/science/article/pii/0029558266908510}

\bibitem{ENDF2018}
Brown D~A, Chadwick M~B, Capote R, Kahler A~C, Trkov A, Herman M~W, Sonzogni
  A~A, Danon Y, Carlson A~D, Dunn M, Smith D~L, Hale G~M, Arbanas G, Arcilla R,
  Bates C~R, Beck B, Becker B, Brown F, Casperson R~J, Conlin J, Cullen D~E,
  Descalle M~A, Firestone R, Gaines T, Guber K~H, Hawari A~I, Holmes J, Johnson
  T~D, Kawano T, Kiedrowski B~C, Koning A~J, Kopecky S, Leal L, Lestone J~P,
  Lubitz C, Damián J~I~M, Mattoon C~M, McCutchan E~A, Mughabghab S, Navr\'atil
  P, Neudecker D, Nobre G~P~A, Noguere G, Paris M, Pigni M, Plompen A,
  Pritychenko B, Pronyaev V, Roubtsov D, Rochman D, Romano P, Schillebeeckx P,
  Simakov S, Sin M, Sirakov I, Sleaford B, Sobes V, Soukhovitskii E~S, Stetcu
  I, Talou P, Thompson I, van~der Marck S, Welser-Sherrill L, Wiarda D, White
  M, Wormald J~L, Wright R~Q, Zerkle M, Žerovnik G and Zhu Y 2018 {\em Nuclear
  Data Sheets\/} {\bf 148} 1 -- 142 ISSN 0090-3752 special Issue on Nuclear
  Reaction Data
  \urlprefix\url{http://www.sciencedirect.com/science/article/pii/S0090375218300206}

\bibitem{Lecluse:17}
Duguet T, Som\`a V, Lecluse S, Barbieri C and Navr\'atil P 2017 {\em Phys. Rev.
  C\/} {\bf 95}(3) 034319
  \urlprefix\url{https://link.aps.org/doi/10.1103/PhysRevC.95.034319}

\bibitem{Nesaraja:16}
Nesaraja C and McCutchan E 2016 {\em Nuclear Data Sheets\/} {\bf 133} 1 -- 220
  ISSN 0090-3752
  \urlprefix\url{http://www.sciencedirect.com/science/article/pii/S0090375216000144}

\bibitem{Note1}
Note that the Lanczos algorithm used the solve for Eq.~(\ref {eq:Sigma_ho})
  does not affect these conclusions since it is specifically designed to
  preserve the strength distribution of response functions~\cite
  {Waldecker:11,Soma:14}.

\bibitem{Soma:14}
Som\`a V, Barbieri C and Duguet T 2014 {\em Phys. Rev. C\/} {\bf 89}(2) 024323
  \urlprefix\url{http://link.aps.org/doi/10.1103/PhysRevC.89.024323}

\end{thebibliography}

\end{document}